%
%
%
%

%
%
\def\ea{{\it et al\/}}
\def\ie{{\it i.e.\ }}
\def\brad{|s_1\rangle}
\def\brai{ \langle s_2|}

\def\dbraki{\langle n_1\,j\,\epsilon|}
\def\dbrakd{|n_2\,j\,\epsilon\rangle}

%
 \newbox\Ancha
 \def\gros#1{{\setbox\Ancha=\hbox{$#1$}
   \kern-.025em\copy\Ancha\kern-\wd\Ancha
   \kern.05em\copy\Ancha\kern-\wd\Ancha
   \kern-.025em\raise.0433em\box\Ancha}}
%
%
\input iopppt
\pptstyle
\jl{2}

\letter{ Recurrence relation for relativistic atomic  matrix elements}

\author{ R P Mart\'inez-y-Romero\dag\footnote{\S}{On sabbatical leave from Facultad de Ciencias, Universidad Nacional Aut\'onoma de M\'exico, e-mail rodolfo@dirac.fciencias.unam.mx  } H N N\'u\~nez-Y\'epez\ddag \footnote{$\|$}{E-mail nyhn@xanum.uam.mx} and A L Salas-Brito\dag \footnote{\P}{E-mail: asb@data.net.mx or asb@hp9000a1.uam.mx }}[R P Mart\'inez-y-Romero \etal\ ]

\address{\dag Laboratorio de Sistemas Din\'amicos, Universidad Aut\'onoma Metropolitana-Azcapotzalco,  Apartado Postal 21-726, C P 04000, Coyoac\'an  D F, M\'exico }
\address{\ddag Departamento de F\'isica, Universidad Aut\'onoma Metropolitana- Iztapalapa, Apartado Postal 55-534, Iztapalapa 09340 D F, M\'exico}

\abs 
Recurrence formulae for arbitrary hydrogenic radial matrix elements are obtained in the Dirac form of relativistic quantum mechanics.  Our approach is inspired on the  relativistic extension of the second hypervirial method that has been succesfully employed  to deduce an analogous relationship in non relativistic quantum mechanics. We obtain first the relativistic extension of the second hypervirial and then the  relativistic recurrence relation. Furthermore, we  use such  relation to  deduce  relativistic versions of the  Pasternack-Sternheimer rule and of the virial theorem.
   \endabs

\submitted 

\date

\pacs{31.30.Jv, 03.65.Pm, 03.20.+i}

 For explaining certain features of atomic spectra, of atom-laser multiphoton transitions and of other atomic or molecular processes, a knowledge of matrix elements of polynomial radial functions between  radial   hydrogenic  eigenstates is obviously of major importance (Moss 1972, Wong and Yeh 1983a, De Lange and Raab 1991,  Quiney \etal\ 1997). Such importance comes about since the $r^\lambda$ terms can be regarded as explicit expressions for interatomic potentials  or  as  terms in a multipolar description of the interaction with the electromagnetic field.
In that area a great deal of work has been done in nonrelativistic quantum mechanics (N\'u\~nez-Y\'epez \etal\ 1997, 1995 and the references therein).  Nowadays, however,  the sophisticated experimental techniques available make necessary the knowledge of such matrix elements between {\sl relativistic} hydrogenic states. Though it is  possible to compute directly such matrix elements (Kobus \ea\ 1987, Wong and Yeh 1983b) the computations quickly become cumbersome thus techniques for evaluating any number of them starting from a few known ones are extremely convenient. For example, motivated by the need of evaluating the curvature-induced modifications of the hydrogen spectrum  Bessis \etal\ (1985) have contributed  in this direction. In spite of this, however, there is not as yet any  valid recurrence relation in relativistic quantum mechanics analogous to the general and useful ones  customarily used  in ordinary quantum mechanics. 

In this work we derive one such recurrence relation between relativistic radial hydrogenic states; this relation expresses the matrix elements of $r^{\lambda }$ and $\beta r^{\lambda}$ (where $\beta$ is the Dirac matrix)  in terms of those of $r^{\lambda-1}$, $\beta r^{\lambda-1}$, $r^{\lambda -2}$, $\beta r^{\lambda-2}$, $r^{\lambda-3}$, and $\beta r^{\lambda -3}$---as given in equations (17) and (20) below. The recurrence relations  we obtain here can be regarded as  a generalization of the non relativistic Blanchard rule (N\'u\~nez-Y\'epez \etal\ 1995, Blanchard 1974) to the  bound hydrogenic radial eigenstates of Dirac relativistic quantum mechanics.  Such relation can be found useful in, for example,  different schemes of perturbation theory used in  atomic calculations (Brack 1983, Dobrovolska and Tutik 1999) or for computing the effect of external fields  on bound electrons (Wong and Yeh 1983a, b).
Please be aware that we only intend to give the main results in this communication, for most of the details and other possible applications see Mart\'inez-y-Romero \etal\ (2000).

The Dirac Hamiltonian, the  Dirac equation for the stationary states of the hydrogen atom, and the corresponding energy eigenvalues,  are

$$ \eqalign{ &H_D= c{\gros\alpha}\cdot {\bf p} + \beta mc^2 - {Z \alpha_F\hbar c\over  r}, \quad  H_D\,\Psi({\bf r})= E\,\Psi({\bf r}), \cr
&E=mc^2  \left( 1+{Z^2\alpha_F^2\over \left(n-j-1/2+\sqrt{(j+1/2)^2-Z^2\alpha_F^2}\right)^2}\right)^{-1/2}  }                     \eqno(1)  $$

\noindent where $\gros\alpha$ and $\beta$ are standard $4 \times 4$ Dirac matrices in the Dirac representation (Bjorken and Drell 1964; Drake 1996, Ch 22 by Grant I P), $Z$ is the atomic number, $r$  the relative distance between the electron and the nucleus, $m$ the electron mass, $c$ the speed of ligth,  $\alpha_F\equiv e^2/4\pi\epsilon_0\hbar c$  the fine structure constant, $e$ the electron charge, $j= 1/2,  3/2,  5/2, \dots$ the total  angular momentum quantum number, and $n=0, 1, 2,\dots$ is the principal quantum number. The obvious rotational symmetry of $H_D$ implies that the bound eigenstates of the hydrogen atom can be written as

$$ 
\Psi_{n, \kappa, m_z}(r,\theta,\phi) = {1\over r}\left( \matrix{F_{n,j,\epsilon}(r)\chi_{\kappa, m_z}(\theta, \phi) \cr\cr \i G_{n, j,\epsilon}(r)\chi_{-\kappa, m_z}(\theta,\phi)}\right). \eqno(2) $$

\noindent where $m_z=-j, -j+1/2, \dots, j-1/2, j$, is the $z$-projection of the total angular momentum quantum number, and $\chi_{\kappa, m_z}(\theta, \phi)$ and $\chi_{-\kappa, m_z}(\theta,\phi)$ are spinor spherical harmonics of opposite parity  and $\kappa=-\epsilon(j+1/2)$ is the eigenvalue of the operator $\Lambda\equiv \beta(1+{\gros \Sigma}\cdot{\bf L})$ (${\gros \Sigma}\equiv{\gros \sigma}\otimes I = \hbox{diag}({\gros\sigma}, {\gros\sigma}$), ${\gros\sigma}$ is the vector with Pauli matrices as components), which can be  seen to commute with $H_D$ (Drake 1996, Ch 22 by Grant I P). We found it convenient to define the quantum number $\epsilon=(-1)^{j+l-1/2}$--- \ie\ $\epsilon$  equals  $+1$ when $l=j+1/2$ or equals $-1$ when $l=j-1/2$--- and use it instead of parity for labeling the eigenstates, so $l=j+\epsilon/2$ and $l'=j-\epsilon/2$, because the small component has the opposite parity to the big one (Mart\'inez-y-Romero \etal\ 1998, 1999). Please notice that another  notation for the spinor functions used in (2) is ${\cal Y}^l_{j, m_z}\equiv \chi_{\kappa, m_z}$ and ${\cal Y}^{l'}_{j, m_z}\equiv \chi_{-\kappa, m_z}$ (Mart\'inez-y-Romero \etal\ 1998, Greiner 1991, Moss 1972). Writing the eigenfunctions in the form (2)  completely solves the angular part of the problem, so we only need  to cope with the radial part of it.  Be aware also that, as we did  in (2),  we prefer to employ  $n$, $j$ and $\epsilon$ rather than just  $n$, and $\kappa$ to label the radial eigenfunctions.  

Let us begin  establishing a relativistic version of the hypervirial result which is known to lead directly---though not straightforwardly because the computations are rather long--- to the Blanchard recurrence relation in nonrelativistic quantum mechanics (N\'u\~nez-Y\'epez \etal\ 1995).  The  radial Hamiltonian associated to  (1) can be obtained using 
 the squared  total (orbital plus spin) angular momentum  ${\bf J}^2={\bf L}^2 + { \gros \sigma} {\gros\cdot}{\bf L} +3/4$, 
and the fact that the operator ${\bf L}^2$, as applied on  eigenstates of the form (2), is equivalent to the action of the operator $j(j+1)+\epsilon \beta (j+1/2) +1/4$ upon the same states. Thus, using $({\gros \alpha}\cdot {\bf r})({\gros \alpha}\cdot {\bf p})=({\gros \Sigma}\cdot {\bf r})({\gros \Sigma}\cdot {\bf p})=  {\bf r}\cdot {\bf p} +\i {\gros \Sigma}\cdot {\bf L}$,  we get
${\gros \alpha}\cdot {\bf p}=\alpha_r\left[p_r -i\beta{(j_k+ {1/ 2})\epsilon_k/ r} \right]$, so we finally obtain  (Mart\'inez-y-Romero \ea\ 2000)

$$ \eqalign{&H_k = \alpha_r\left[p_r -i\beta{\epsilon_k\over r} \left(j_k+ {1\over 2}\right)\right] +\beta m + V(r), \cr\cr         
&H_k\psi_k(r)= E_k\psi_k(r),} \eqno(3) $$

\noindent where we use units such that $\hbar=c=1$,  $V(r)$ is an arbitrary radial potential, $k$ is a label introduced  for the sake of later convenience, and

$$ \eqalign{&\alpha_r= {\gros \alpha}\cdot {{\bf r}\over r}=\pmatrix{0& -1\cr -1& 0}, \qquad p_r=-{\i \over r} {d\over dr} \,r ,\cr\cr
&\psi_k(r)\equiv \psi_{n_k,j_k,\epsilon_k}(r)= {1\over r}\pmatrix{F_{n_k,j_k,\epsilon_k}(r)\cr \i G_{n_k,j_k,\epsilon_k}(r)}.} \eqno(4) $$

\noindent  As it should be clear from the matrix expression for $\alpha_r$ in the previous equation, we can use a 2$\times$2 representation valid for the radial eigenstates given in (4)---where the now $2\times 2$  $\beta$-matrix is just diag$(1,-1)$ with numerical entries (Constantinescu and Magyari 1971, p 382). Using this  representation the purely radial Dirac equation reduces to

$$ \fl \left[\matrix{m+(V(r)-E)& Z{\epsilon(j+1/2)/ r} -{d/ dr}\cr\cr
          Z\epsilon(j+1/2)/r +d/dr& m-(V(r)-E)}\right] \left[\matrix{F_{nj\epsilon}(r)\cr\cr G_{nj\epsilon}(r)}\right]=0; \eqno(5)
                       $$

\noindent  we have to point out, however, that using this representation is not strictly necessary and that all our results are representation independent.

The key relationship needed for deriving the relativistic recurrence relation stems directly from  equation (3)---compare with equation (2) and the nonrelativistic discussion that follows in (N\'u\~nez-Y\'epez \etal\ 1995 pp. L526--L527). 

Let us first compute  matrix elements of the radial function $\xi(r)\equiv H_2f(r)-f(r)H_1$ between radial eigenstates of the hydrogen atom in the energy basis, where  $f(r)$ is an arbitrary radial function and the $H_i$ are the radial Hamiltonians appearing in (3). Evaluating such radial matrix elements we get 

$$\eqalign{ (E_2 - E_1) \langle &n_2\,j_2\,\epsilon_2|f(r)|n_1\,j_1\,\epsilon_1\rangle\cr &  = \langle n_2\,j_2\,\epsilon_2 |H_2f(r)- f(r)H_1|n_1\,j_1\,\epsilon_1\rangle\cr& = -\i \langle n_2\,j_2\,\epsilon_2|\alpha_r\left(f'(r)+ {\Delta^-_{21}\over 2r}\beta f (r)\right)|n_1\,j_1\,\epsilon_1\rangle,\cr
\hbox{where} \quad |n\,j\,\epsilon&\rangle \equiv \pmatrix{F_{nj\epsilon}\cr \i G_{nj\epsilon}},
}\eqno(6)$$

\noindent  we have defined  the quantities $ \Delta^{\pm}_{21} \equiv  \epsilon_2(2j_2 + 1) \pm \epsilon_1(2j_1 + 1)$, and the expressions used for the $\langle n_2\,j_2\,\epsilon_2|f(r)|n_1\,j_1\,\epsilon_1\rangle$ and the $\langle n_2\,j_2\,\epsilon_2|\beta f(r)|n_1\,j_1\,\epsilon_1\rangle$ matrix elements are 

$$\eqalign{ \langle n_2 j_2 \epsilon_2| f(r)|n_1 j_1 \epsilon_1\rangle  &= \int f(r) \left[ F_1(r)F^*_2(r) + G_1(r)G^*_2(r) \right ] dr,\cr
   \langle n_2 j_2 \epsilon_2|\beta f(r)|n_1 j_1 \epsilon_1\rangle & = \int f(r) \left[ F_1(r)F^*_2(r) - G_1(r)G^*_2(r) \right ]dr.}\eqno(7)$$ 

\noindent We then apply  the result (6) to the  function $H_2\xi(r)- \xi(r) H_1 $, to obtain

$$\eqalign{ &(E_2-E_1)^2\brai  f(r)   \brad= \cr &\brai-{\Delta^{-}_{21} \over 2r^2} \beta f(r)-f''(r)-{\Delta^{-}_{21} \over 2r} f'(r) \beta- {\Delta^{-}_{21} \over r} f(r) \beta{d\over dr} + \cr &{\Delta_{21}^{+}\over 2r}  f'(r)\beta  +  \left({\Delta^{-}_{21}\over 2r}\right)^2 f(r) + 2\i \alpha_r\beta m\left(f'(r) + {\Delta^{-}_{21} \over 2r}\beta f(r)\right)\brad ;}\eqno(8)  $$
\par

\noindent where the subscripts in the radial components, $F$ and $G$,  stand  for the three quantum numbers $n$, $j$, $\epsilon$, and we have assumed $ \Delta^-_{21}\neq 0$. From now on, as we did in equation (8), we use the {\sl shorthand} $|s_i\rangle\equiv |n_i\,j_i\,\epsilon_i\rangle$ for the states of the system. It is to be noted that (8) can be regarded as the relativistic equivalent to the hypervirial obtained in (N\'u\~nez-Y\'epez \etal\ 1995, equation (8), p.\ L526).

 However, in the Dirac case we are dealing with, at difference to what happens in the nonrelativistic case, equation (8) is not enough for deriving the recurrence relation between matrix elements of powers of $r$, we also need the following results (Mart\'inez-y-Romero \etal\ 2000):

\noindent a) A second order iteration for certain non diagonal matrix elements, where we  use for computing the elements, the radial function $H_2\xi(r)+\xi(r) H_1$  

$$  \eqalign{  (E_2 ^2 &- E_1^2)\brai f(r) \brad = \brai H_2\xi + \xi H_1\brad = \cr
&\brai  -{2f'(r)\over r}  + {\Delta^-_{21} \over 2r^2}\beta f(r) - f''(r) -2f'(r) {d\over dr}  + \cr &{\Delta_{21}^+ \Delta^-_{21} \over 4r^2} f(r) -  2\i \alpha_r \left(f'(r) + {\Delta^-_{21} \over 2r}\beta f(r)\right)V(r)  \brad,   } \eqno(9)$$

\noindent b) the following  matrix elements 

$$\eqalign{ -\i (E_2 + E_1)\brai \alpha_r f(r)\brad= 
-\brai &{2f(r)\over r}  + f'(r) +2 f(r) {d\over dr}  \cr -&{\Delta^-_{21} \over 2r}\beta f(r) + 2\i\alpha_r V(r) f(r)\brad, }  \eqno(10) $$ 

\noindent c)  the matrix elements of the radial function $-\i (H_2\alpha_rf(r)-\alpha_rf(r)H_1)$,  lead to 

$$ \eqalign{ -\i(E_2 - E_1)\brai &\alpha_r f(r)\brad = \cr
 &\brai  -f'(r) + {\Delta_{21}^+ \over 2r} \beta f(r) + 2\i\alpha_r\beta m f(r) \brad.}\eqno(11) $$

\noindent and, d) the matrix elements of the radial function $H_2\beta f(r)+\beta f(r)H_1$, lead to (Mart\'inez-y-Romero 2000)

$$ \eqalign{(E_2 + E_1)\brai \beta f(r)\brad = &
\brai \i\beta\alpha_r f'(r) -\i \alpha_r {\Delta^-_{21} \over 2r} f(r)\cr &+ 2\left[m + \beta V(r)\right]f(r)\brad.} \eqno(12) $$

\noindent Equations (6) and (8--12) are the basic equations for our problem. For more details on the computations please see Mart\'inez-y-Romero \etal (2000). \par

Up to this point, our results are valid both for an arbitrary radial potential $V(r)$ and for an arbitrary radial function $f(r)$ but, to be specific, let us assume that $V(r)$ is precisely the Coulomb potential, \ie\

$$ V(r)= -{Z\over r},  \eqno(13)  $$ 

\noindent where we have taken $e^2/4\pi \epsilon_0=1$. In this work we  consider only the case of functions of the form $f(r)=r^\lambda$; so,  putting $f(r)=r^{\lambda-1}$ in equation (6) and  in equation (10), extracting the term with $-\i\alpha_r \Delta^-_{21}\beta r^{\lambda-2}$ from the former  and the term with $-2\lambda r^{\lambda-1}d/dr $ from the latter and substituting them into equation (9), but evaluated using $f(r)=r^\lambda$, we get

$$ \eqalign{&(E_2^2 - E_1^2)\brai r^\lambda\brad =\brai {\Delta^-_{21}\Delta_{21}^+ \over 4} r^{\lambda -2}  + {\Delta^-_{21}\over 2} (1-\lambda) \beta r^{\lambda -2} +\cr &
Z\left[2\i\alpha_rr^{\lambda -2}  (1-\lambda) - 2(E_2 - E_1) r^{\lambda -1} \right] - (E_2 + E_1) \lambda \i\alpha_r r^{\lambda -1}\brad.} \eqno(14) $$ 

\noindent   We next use $f(r)=r^{\lambda}$ in equation (6),  extract the term with $2\i \alpha_r \beta m r^{\lambda-1}$ and substitute it into (11) but evaluated using $f(r)=r^{\lambda-1}$, to obtain the following result

$$\eqalign{& \left[ (E_2 - E_1) -{4m\lambda\over \Delta^-_{21}}\right]\brai(-\i\alpha_rr^{\lambda -1})\brad=\cr
&\brai-(\lambda -1 ) r^{\lambda -2}  - {4m\over \Delta^-_{21} }(E_2-E_1)r^\lambda  + {\Delta^+_{21}\over 2}\beta r^{\lambda -2}\brad; }\eqno(15)$$

\noindent then, we use $f(r)=r^\lambda$ in equation (12) and  follow the same procedure as above, to get instead

$$\eqalign{ &\left[ (E_2 - E_1) - {\Delta^-_{21} m\over \lambda} \right]\brai (-\i\alpha_r r^{\lambda -1})\brad =\brai -(\lambda -1)r^{\lambda -2} +\cr& {4m^2\over \lambda} r^\lambda +{\Delta^+_{21}\over 2} \beta r^{\lambda -2}  
 - {4Z m \over \lambda }\beta r^{\lambda -1} -{2m\over\lambda }(E_2 + E_1) \beta r^{\lambda} \brad. }\eqno(16) $$

\noindent To continue, we evaluate equation (15) first using $f(r)=r^{\lambda-1}$ and then  using  $r^{\lambda-2}$, next we extract the term $-\i\alpha_r\lambda (E_2+E_1) r^{\lambda-1}$ from the former and the term $-2\i Z \alpha_r (\lambda-1) r^{\lambda-2}$ from the latter, to finally obtain, on substituting these extracted terms  into (14),  the recurrence relation

$$ \eqalign{ c_0 \brai r^\lambda \brad =\sum_{i=1}^{3} c_i\brai r^{\lambda -i} \brad + \sum_{i=2}^{3} d_i\brai \beta r^{\lambda -i}\brad, } \eqno(17) $$

\noindent where the numbers $c_i$,  $i=0,\dots 3$ are given by

$$\eqalign{c_0 & = {(E_2^2 - E_1^2)(E_2 - E_1)\Delta^-_{21}\over (E_2 - E_1)\Delta_{21}^- - 4m\lambda}, \cr
c_1 & = -{2 Z (E_2 - E_1)^2 \Delta_{21}^-\over (E_2 - E_1)\Delta_{21}^- - 4m(\lambda -1)},\cr
c_2 & = {\Delta_{21}^-\Delta_{21}^+\over 4} -\lambda(\lambda -1){(E_1 + E_2)\Delta_{21}^-\over (E_2 - E_1)\Delta_{21}^- -4m\lambda},\cr
c_3& = {-2 Z(\lambda -1)(\lambda -2)\Delta_{21}^-  \over (E_2 - E_1) \Delta_{21}^- -4m(\lambda -1) },} \eqno(18)$$

\noindent and the numbers $d_i$, $i=2$ and 3, by
 
$$ \eqalign{
d_2 & = {\Delta_{21}^-\over 2} \left[     (1-\lambda) + {\lambda (E_2 + E_1)\Delta_{21}^+\over(E_2- E_1) \Delta_{21}^- -4m\lambda}\right],\cr
d_3 & = {Z (\lambda -1) \Delta_{21}^- \Delta_{21}^+ \over(E_2 - E_1) \Delta_{21}^- - 4m
(\lambda -1)}.}\eqno(19) $$

\noindent Equation  (17), together with the specific values for the $c_a$ and the $d_a$, can be regarded as the direct relativistic version of the Blanchard (1974) relation ---compare with equations (10) and (11) in N\'u\~nez-Y\'epez \etal\ (1995). These relations are valid inasmuch  as  the number $ w_1+w_2+\lambda+1$ is greater than zero, where $w_i\equiv\sqrt{(j_i+1/2)^2-Z^2\alpha_F^2}=\sqrt{\kappa_i^2-Z^2\alpha_F^2}$.  This condition amounts basically to the requirement that any integrand should be at least of the form $1/r^{1+\gamma}$ with $\gamma >0$  (Mart\'inez-y-Romero \etal\ 2000).

 The extra complication in the recurrence relation (17), that does not occur in the nonrelativistic case, is the explicit appearance of matrix elements of powers of $r$ times the $\beta$ matrix. In a way, this is just a matter of a sign change in an integral---as it should be obvious from equation (7).  However, since  it is not possible to avoid the $\beta$ dependence  in  (17), as it stands such equation  does not really allow the computation of $\brai r^\lambda \brad$ in terms of the $\brai r^{\lambda-i} \brad\quad i=1,2,3$ as it undeniably happens in nonrelativistic quantum mechanics (N\'u\~nez-Y\'epez \etal\ 1995), something additional is needed. 

For obtaining the extra  information required,  we only need to multiply equation (16) times $(E_2-E_1)-4m \lambda /\Delta^-_{21} $ and equation (15) times $ (E_2-E_1)- m  \Delta^-_{21}/\lambda $, and then  substract the results,  to finally obtain the lacking recurrence relation for  the matrix elements involving the $\beta$ matrix times  $r$-powers, namely

$$  \eqalign{e_0  \brai \beta r^\lambda \brad = &b_0 \brai r^{\lambda}\brad + b_2 \brai r^{\lambda-2}\brad + e_1 \brai \beta r^{\lambda-1}\brad\cr & + e_2 \brai \beta r^{\lambda-2}\brad,} \eqno(20) $$

\noindent where the numbers $d_i$ and $e_i$ $i=1, 2, 3$ are given by

$$ \eqalign{b_0=& 4\lambda\left[(E_2-E_1)^2 -4 m^2 \right], \cr
             b_2=&(1-\lambda)\left[(\Delta_{21}^{-})^2-4\lambda^2\right], \cr
             e_0=&2(E_2+E_1)[(E_2-E_1)\Delta^-_{21}-4m\lambda],\cr
             e_1=&4Z[4m\lambda-(E_2-E_1)\Delta^-_{21}],\cr
             e_2=& {\Delta_{21}^+\over 2}[(\Delta_{21}^{-})^2-4\lambda^2].} \eqno(21) $$

\noindent The validity conditions of this recurrence relation is that the number $ (w_1+w_2+\lambda+1)$ be greater than zero, exactly as before.

 Equations (17) and (20) are together the useful recurrence relations for evaluating radial matrix elements between relativistic radial hydrogenic states; in this sense  they are thus the actual relativistic generalization of Blanchard rule.

The recurrence relation obtained above (17) pressupose $\Delta^-_{21}\neq 0$, but, for studying the diagonal case, we must have $\epsilon_1= \epsilon_2$ and $j_1=j_2$ (in other words $\kappa_1=\kappa_2$) which  precisely imply that  $\Delta^{-}_{21}= 0$. In such instance we cannot apply that relation,  we have  to rederive the recurrence relation using as  starting  equations 

$$(E_2 -E_1)\langle n_1\, j\,\epsilon | f(r) \dbrakd = \dbraki(-\i\alpha_rf'(r))\dbrakd,\eqno(22)$$ 

$$\eqalign{(E_2 - E_1) \dbraki-\i\alpha_r f(r)\dbrakd =&\dbraki-f'(r)  + {\Delta^+\over 2r}\beta f(r) +\cr &+ 2i \alpha_r \beta m f(r)\dbrakd ,} \eqno(23)$$

\noindent and 

$$\eqalign{ (E_2 + E_1)\dbraki\beta f(r)\dbrakd &=\dbraki-\i\alpha_r \beta f'(r) \cr &+ 2\left(m + \beta V(r)\right)f(r)\dbrakd,} \eqno(24) $$

\noindent  where,  for avoiding possible misunderstandings, we have reverted to  explicitly writing the quantum numbers in the states and $\Delta^+\equiv 2\epsilon(2j + 1).$ To get the result we want, we evaluate equation (22) with $f(r)=r^\lambda,$ then we put $f(r)=r^{\lambda -1}$ in equation (23), and  $f(r)=r^\lambda$ in equation  (24); thence,  using essentially a procedure similar to the used in the $\Delta_{21}^- \neq 0$ case outlined above, we  finally obtain (Mart\'inez-y-Romero \etal\ 2000)

$$ \eqalign{ &\left[ (E_2 - E_1)^2 - 4m^2 \right]  \dbraki r^\lambda \dbrakd = \lambda {\Delta_{21}^+\over 2}\dbraki \beta r^{\lambda-2}\dbrakd \cr & -4m\dbraki\beta r^{\lambda-1}\dbrakd-2m(E_2+E_1)\dbraki\beta r^{\lambda}\dbrakd\cr& -\lambda(\lambda-1) \dbraki r^{\lambda-2}\dbrakd.} \eqno(25) $$

\noindent This is the version of the recurrence relation (17) which is valid when $\Delta_{21}^-=0$. On the other hand, relation (20)  can be  directly written in the case $\Delta_{21}^-=0$, giving a equation entirely equivalent to (25); so, in the case where $\Delta_{21}= 0,$ we have relation (25) and relation (20) with the restriction $\Delta_{21}=0$. 

As happens with the non relativistic recurrence relation (N\'u\~nez-Y\'epez \ea\ 1995), various of its particular cases  have interest on their own; for example  when $\lambda=0$, from (25) we directly obtain

$$\eqalign{\left[ (E_2 - E_1)^2 - 4m^2 \right]\delta_{n_1 n_2} &=- 4m\, \dbraki {\beta \over r}\dbrakd\cr & -2m\,(E_2 + E_1)\, \dbraki\beta\dbrakd,}\eqno(26)$$

\noindent where $\delta_{ij}$ is a Kronecker delta.  This relation (26) could be regarded as a relativistic version of the well-known Pasternak-Sternheimer (1962) rule of non relativistic quantum mechanics, which says that the expectation   value between hydrogenic states of the $1/r^2$ potential,  vanishes when  the orbital angular momenta of the  states $1$ and $2$ coincide, \ie\ when $l_1=l_2.$   We remark that in the relativistic case the expectation value of the  $1/r$ potential (which corresponds to the square root of $1/r^2$)  times $\beta$,  does {\bf not} necessarily vanish even when the total angular momenta of the two states coincide: $j_1=j_2$.  This agrees with the known non relativistic fact that the Pasternack-Sternheimer rule is applicable to eigenfunctions of potentials whose energy eigenvalues depend only on the principal quantum number---which is not the case for the hydrogen atom in Dirac relativistic quantum mechanics.\par

 Furthermore, in the completely diagonal case (\ie when $n_1=n_2$, $j_1=j_2$, and $\epsilon_1=\epsilon_2$; or just $n_1=n_2$ and $\kappa_1=\kappa_2$) we easily find ($\langle O\rangle$ stand for $O$'s expectation value)

$$  m=-\left< \beta V(r)\right> +  E\,  \left<  \beta \right> =Z\left<{\beta\over r}\right>
 + E\,  \left<  \beta \right> ,
 \eqno(27)$$

\noindent or, using $\langle \beta \rangle=E/m $ (De Lange and Raab 1991),

$$ E^2=m\, \left< \beta V(r)\right> +m^2 =-m\,Z \left<  {\beta\over r}\right> +m^2 . \eqno(28)  $$

\noindent This last expression can be regarded as  another version of the relativistic virial theorem in a form different from the one proposed by  Kim (1967). Please notice that the first equalities in equations (27) and (28) above are valid for an arbitrary central potential $V(r)$. 

 Some other useful  things that can be obtained  from both the relativistic hypervirial [equation (8)] and the  subsequent results [equations (9--12)], and from the relativistic recurrence relation presented in this contribution [equations (17) and (20)]; but such discussions as well as the many details left out from our presentation, will appear in a more detailed  article (Mart\'inez-y-Romero \etal\ 2000).  Let us pinpoint again that for our results to have meaning we need that the exponent in the recurrence relations, $\lambda$, comply with $ (\lambda+1) >-w_1-w_2$, where we assume that the $w_i=\sqrt{(j_i+1/2)^2-Z^2\alpha_F^2}$ are real numbers (Mart\'inez-y-Romero \etal\ 2000). This condition has a similar form to the one required by Blanchard (1974) in the non relativistic case.

We have just learnt of the plans to test CPT and Lorentz invariance studying the 1s-2s two photon transition in hydrogen, in which the recurrence relations proposed here can be of some help for the computations (Bluhm \etal 2000).

\ack  We acknowledge with thanks the collaboration of C Cisneros and the  comments of  J J Pe\~na and V Gaftoi. Last but not least, we want to thank the friendly support of M Dochi, C F Quimo, U Kim, Q Chiornaya, C Sabi, M Sieriy, Ch Cori,  S Mahui, R Sammi, M Becu, K Zura, and M Mati. This paper is dedicated to the memory of C Ch  Ujaya, F Cucho, R Micifus, and B Ch Caro. \par

\references
\refjl {Bessis N, Bessis G, and Roux D 1985} {Phys. Rev. A } {32} {2044}
\refbk  {Bjorken J D and  Drell S 1964}  {Relativistic Quantum Mechanics} {(New York: Mac Graw-Hill)}
\refjl {Bluhm R, Kosteleck\'y V A, and Russell N 2000} {arXiv:hep-ph/0003223}{}{}
\refjl{Brack M 1983} {Phys. Rev. D } {27} {1950}  
\refjl {Blanchard P 1974} {J. Phys. B: At. Mol. Opt. Phys.} {7} {1993}
\refbk{Constantinescu F and Magyari E 1971} {Problems in Quantum Mechanics} {(Oxford: Pergamon Press)} 
\refbk{De Lange O L and Raab R E 1991} { Operator Methods in Quantum Mechanics} {(Oxford: Clarendon)}
\refjl{Dobrovolska I V and Tutik R S 1999} {Phys. Lett. A } {260} {10}
\refbk {Drake G W F (Ed) 1996} {Atomic, Molecular and Optical Physics Handbook} {(Woodbury: American Institute of Physics) Ch 22} 
\refbk{Greiner W 1991} { Theoretical Physics 3: Relativistic quantum 
mechanics} {(Berlin: Springer)} 

\refjl {Kim Y-K 1967} {Phys. Rev.} {154} {17}

\refjl {Kobus J, Karkwowski J and Jask\'olski W 1987} {J. Phys. A: Math. Gen. } {20} {3347}
\refjl {Mart\'inez-y-Romero R P,  Salda\~na-Vega J and Salas-Brito A L 1998} {J. Phys. A: Math. Gen.} {31} {L157}  
\refjl {Mart\'inez-y-Romero R P,  Salda\~na-Vega J and Salas-Brito A L 1999} {J.  Math. Phys.} {40} {2324}  
\refjl {Mart\'inez-y-Romero R P,  N\'u\~nez-Y\'epez H N and Salas-Brito A L 2000} {Phys. Rev. A} {submitted} {} 
\refbk {Moss R E 1972} {Advanced Molecular Quantum Mechanics} {(London: Chapman and Hall)}
\refjl{N\'u\~nez-Y\'epez H N,  L\'opez J  and Salas-Brito A L 1995} {J. Phys. B: At. Mol. Opt. Phys.} {28} {L525}
\refjl {N\'u\~nez-Y\'epez H N, L\'opez J, Navarrete D and Salas-Brito A L 1997} {Int. J. Quantum Chem.} {62} {177} 
\refjl{Pasternack S and Sternheimer R M 1962} {J. Math. Phys.} {3} {1280}

\refjl {Quiney H M, Skaane H and Grant I P 1997} {J. Phys. B: At. Mol. Opt. Phys.} {30} {L829}

\refjl{Wong M K F and Yeh H-Y 1983a} {Phys. Rev. A}  {27} {2300}

\refjl{Wong M K F and Yeh H-Y 1983b} {Phys. Rev. A}  {27} {2305}

 \vfill
 \eject
  \end